\documentclass[preprint,12pt]{elsarticle}
\journal{Nuclear Physics B}

\usepackage[T1]{fontenc} 

\usepackage{url}
\usepackage{graphicx}

\usepackage{latexsym}
\usepackage{amssymb}
\usepackage{amsfonts}

\usepackage{amsmath}
\usepackage[caption=false]{subfig}

\usepackage{breqn}

\def\dsqcup{\sqcup\mathchoice{\mkern-7mu}{\mkern-7mu}{\mkern-3.2mu}{\mkern-3.8mu}\sqcup}

\newcommand{\bea}{\begin{eqnarray}}
\newcommand{\eea}{\end{eqnarray}}
\newcommand{\be}{\begin{equation}}
\newcommand{\ee}{\end{equation}}

\newcommand{\dd}{{\rm d}}
\newcommand{\nn}{\nonumber}

\allowdisplaybreaks[4]
 
\begin{document}
\begin{frontmatter}

\title{
\normalfont
Evaluating Multiple Polylogarithm Values at Sixth Roots of Unity up to Weight Six}

\author[Mainz]{J.M. Henn}
\ead{henn@uni-mainz.de}

\author[SRCC]{A.V.~Smirnov}
\ead{asmirnov80@gmail.com}

\author[SINP]{V.A.~Smirnov\corref{cor1}}
\ead{smirnov@theory.sinp.msu.ru}

\cortext[cor1]{Corresponding author}

\address[Mainz]{
Institut f\"{u}r Physik,
Johannes Gutenberg-Universit\"{a}t Mainz, 55099 \\ Mainz, Germany}
\address[SRCC]{Research Computing Center, Moscow State University, 119992\\ Moscow, Russia}
\address[SINP]{Skobeltsyn Institute of Nuclear Physics of Moscow State University,  119992 \\ Moscow, Russia}

\begin{abstract}
We evaluate multiple polylogarithm values at sixth roots of unity up to weight six,
i.e. of the form $G(a_1,\ldots,a_w;1)$
where the indices $a_i$ are equal to zero or a sixth root of unity, with $a_1\neq 1$. 
For  $w\leq 6$, we construct bases of the linear spaces generated by 
the real and imaginary parts of $G(a_1,\ldots,a_w;1)$
and obtain a table for expressing  
them as linear combinations of the elements of the bases. 
\end{abstract}
\begin{keyword}
Multiloop Feynman integrals \sep multiple polylogarithms \sep sixth roots of unity \sep shuffle and stuffle relations
\end{keyword}
\end{frontmatter}
\newpage
%

\section{Introduction}

In the study of Feynman integrals multiple polylogarithms (MPL), also called 
hyperlogarithms \cite{Lappo-Danilevsky} or Goncharov polylogarithms \cite{Goncharov:1998kja}, 
play an important role. 
They are defined as iterated
integrals over integration kernels d$t/(t-a_{i})$, for some set of numbers $a_{i}$.
More precisely,
\be\label{eq:Mult_PolyLog_def}
 G(a_1,\ldots,a_w;z)=\,\int_0^z\frac{1}{t-a_1}\,G(a_2,\ldots,a_w;t)\, \dd t
\ee
with $a_i, z\in \mathbb{C}$ and $G(z)=1$. 

In the special case where  $a_i=0$ for all $i$, the corresponding integral is divergent 
and instead one defines
\be
G(0,\ldots,0;z) = \frac{1}{n!}\,\log^n z \;.
\ee
If $a_w\neq 0$ and $\rho\neq 0$, then 
$G(\rho a_1,\ldots,\rho a_w;\rho z) = G(a_1,\ldots,a_w;z) $ so that one can express
such MPL in terms of $G(\dots ;1)$.
The length $w$ of the index vector is called the weight. A very well known and studied subset
of MPL are harmonic polylogarithms~\cite{Remiddi:1999ew}, i.e. MPL with the letters $\{0,-1,1\}$.

MPL are very important functions in the field of evaluating multiloop Feynman integrals
because they appear quite naturally and are very often involved in results -- see, e.g., recent papers
\cite{Henn:2013woa,Henn:2014lfa,Caola:2014lpa,Gehrmann:2014bfa,vonManteuffel:2014mva,Dulat:2014mda}.

One often needs to evaluate Feynman integrals at special values of some variable corresponding to a physically 
interesting special value of the Feynman integrals. For example, in the case of massive form factor integrals, 
a specific value of such a variable corresponds to the threshold of creating a pair of massive particles. 
This leads to MPL with a certain constant set of indices.
In this paper, we focus on the case where 
the indices $a_i$ are equal either to zero or a sixth root of unity, i.e.
taken from the seven-letters alphabet $\{0,r_1,r_3,-1,r_4,r_2,1 \}$ with
\be
r_{1,2}=\frac{1}{2}\left(1\pm\sqrt{3}\,{\rm i} \right)=\lambda^{\pm 1}\;, \;\;\;
r_{3,4}=\frac{1}{2}\left(-1\pm\sqrt{3}\,{\rm i} \right)=\lambda^{\pm 2}\;,\;\;\;
\lambda=e^{\pi {\rm i}/3}=r_1\;.
\label{letters}
\ee
This specific case appears in a number of physically interesting problems, e.g. 
in 3-loop QCD corrections to the rho-parameter \cite{Broadhurst:1998rz},
in  the $H \to Z \gamma$ decay at two loops \cite{Bonciani:2015eua}.
Other related references where sixth roots of unity appear include
\cite{Panzer:2016snt,Kalmykov:2010xv,Ablinger:2011te,Bloch:2013tra,Bloch:2014qca}.

Our project was motivated by the necessity to significantly simplify results of
the evaluation of certain massive form factors at three loops \cite{Henn:2016tyf}
and the corresponding vertex master integrals \cite{Henn:2016kjz}, both at general momentum squared and 
at two-particle threshold. These results were indeed written in terms of the constants of the form $G(a_1,\ldots,a_w;1)$ up to $w=6$,
with the indices from the above seven-letters alphabet. 
As we will show below, these constants can be expressed in terms of elements
of the corresponding bases, i.e. irreducible constants.
 
On the other hand, studying relations between special values of
MPL is an interesting mathematical problem -- see, e.g., 
\cite{Goncharov:1998kja,Broadhurst:1998rz,Panzer:2016snt,
Ablinger:2011te,Deligne,Broadhurst:1996az,Broadhurst:2014jda,Brown:2011ik,Zhao,Zagier,Broadhurst_lecture,DG}.
As we will see below, our results provide interesting interconnections with various
mathematical results and conjectures.

In the next section, we specify the goal of this paper, and mention previous related results in the literature.
In Section~3, we solve various relations for the real and imaginary parts of $G(a_1,\ldots,a_{w};1)$ 
up to $w=6$ recursively with respect to $w$.
We construct bases of the linear spaces generated by 
Re$\,G(a_1,\ldots,a_w;1)$ and Im$\,G(a_1,\ldots,a_w;1)$
and explain how the table of results expressing 
them as linear combinations of the elements of the bases was obtained. 
In Conclusion,
we discuss our results and perspectives. Elements of our bases up to weight 6 are described in the Appendix.

\section{Preliminaries}
   
We consider MPL  $G(a_1,\ldots,a_w;1)$ up to $w=6$, 
with indices $a_i$ equal either to zero or a sixth root of unity, i.e.
taken from the seven-letters alphabet $\{0,r_1,r_3,-1,r_4,r_2,1 \}$ of eq. (\ref{letters}).
We imply that $a_1\neq 1$, i.e. we consider convergent MPL.
Our goal is to express any MPL from this set as a linear combination of some irreducible elements.
 
At a given weight $w$, let us denote by $M(w)$ the linear space composed of linear combinations
of $6\times 7^{w-1}$ numbers $G(a_1,\ldots,a_w;1)$ (with $a_1\neq 1$) with rational coefficients.
One can consider separately the real and imaginary parts of the MPL
\be
G(a_1,\ldots,a_{w};1)= G_R(a_1,\ldots,a_{w})+ {\rm i}\, G_I(a_1,\ldots,a_{w})
\label{reim}
\ee
and define similarly the linear spaces $M_R(w)$ and $M_I(w)$.
In particular, products of elements $M_I(w')$ and $M_R(w'')$ of lower weight 
are elements of the spaces $M_R(w)$ and $M_I(w)$ (depending on their parity)
due to the product formulas (\ref{shuffle}) and (\ref{stuffle}) below.

It is well known that MPL satisfy various relations. 
Using shuffle, stuffle relations and some relations following from certain changes of variables,
Broadhurst constructed~\cite{Broadhurst:1998rz} such bases up to weight~3, where all the elements
are described in terms of Cl$_2(\pi/3)$, $\log(2)$, $\log(3)$, $\pi$, $\zeta(3)$ and Li$_2$ and Li$_3$ of certain arguments --
see Appendix where our bases up to weight~3 are explicitly expressed in terms of these constants.
Bases of weight~4 can be found in \cite{FrancescoMoriellosthesis}.
Bases for the  alphabet with letters $0,1,-1$ were constructed in~\cite{Broadhurst:1996az}. (See also
\cite{Blumlein:2009cf}.)
Bases for the multiple Deligne values, i.e. for the alphabet with letters $0,1,r_1$ were presented in \cite{Broadhurst:2014jda}.
Some constants present in results for Feynman integrals up to weight 5 were discussed in
\cite{Kalmykov:2010xv,FrancescoMoriellosthesis,Fleischer:1999mp,Davydychev:2000na}.
There is also a recent method based on the motivic coaction \cite{Brown:2011ik} which can be used to construct bases.
In this paper, we will us the method of \cite{Broadhurst:2014jda}, in the form presented in ref.  \cite{Zhao}.


\section{Solving relations for MPL}

The MPL can be represented as multiple nested sums
\bea
\textrm{Li}_{m_1,\ldots,m_k}(x_1,\ldots,x_k) = 
\sum_{n_k=1}^\infty\sum_{n_{k-1}=1}^{n_k-1}\ldots\sum_{n_1=1}^{n_{2}-1}
{x_1^{n_1}\over n_1^{m_1}}\ldots{x_k^{n_k}\over n_k^{m_k}}
&& \nn \\  && \hspace*{-80mm}
=(-1)^k\,G\left(\underbrace{0,\ldots,0}_{m_k-1},\frac{1}{x_k}, \ldots, 
\underbrace{0,\ldots,0}_{m_1-1},\frac{1}{x_1 \ldots x_k};1\right)\,.
 \label{eq:Lim_def}
\eea
Since the arguments of the Li-functions involved have the form
$x_i=\lambda^{p_i}$ for $p_i=0,\ldots,5$ we introduce an auxiliary function
\be
L_{m_1,\ldots,m_k}(p_1,\ldots,p_k) =
\textrm{Li}_{m_1,\ldots,m_k}\left(\lambda^{p_1},\ldots,\lambda^{p_k}\right)\;.
\label{Lfunction}
\ee
 
To achieve our goal, we start with using the so-called shuffle, stuffle, regularization and distribution relations.
We do not exemplify these relations because they are well known. Various examples of these relations in lower
weights can be found, e.g., in~\cite{FrancescoMoriellosthesis}.

The shuffle relations have the form
\be
G(a_1,\ldots,a_{w_1};z)\,G(b_1,\ldots,b_{w_2};z)=\sum_{c\in a \dsqcup b}\,G(c_1,\ldots,c_{w_1+w_2};z)\;,
\label{shuffle}
\ee
where 
$a_1 \neq 1$ and $b_1 \neq 1$ and 
the sum is over the set of all the mergings of the vectors $a=(a_1,\ldots,a_{w_1})$ and $b=(b_1,\ldots,b_{w_2})$, i.e. 
all possible concatenations in which relative orderings of $a$ and $b$ are preserved. 
These relations follow from the integral representation (\ref{eq:Mult_PolyLog_def}).
We will apply them at $z=1$. 
 
The so-called stuffle relations follow from the sum representation.
They are written down in the language of the
functions (\ref{Lfunction}),
\bea
L_{a_1,\ldots,a_{w_1}}(p_1,\ldots,p_{w_1})\, L_{b_1,\ldots,b_{w_2}}(q_1,\ldots,q_{w_2})=
\sum_{c\in a*b}\,L_{c_1,\ldots,c_{n}}(r_1,\ldots,r_{n})\;,
\label{stuffle}
\eea
where the asterisk denotes the stuffle product.
In contrast to the shuffle product, the sum goes, in particular, over all
the shuffles of the
vectors $a=(a_1,\ldots,a_{w_1})$ and $b=(b_1,\ldots,b_{w_2})$ where the
shuffling of
$(p_1,\ldots,p_{w_1})$ and $(q_1,\ldots,q_{w_2})$
resulting in a vector $(r_1,\ldots,r_{n})$ is the same as for $a$ and $b$.
In addition, some of the letters from the sets $a$ and $b$ are allowed to
be put together and, in this case,
the indices $a_i$ and $b_j$ are summed up and the corresponding arguments
$p_i$ and $q_j$ are summed
up modulo six.  Relations (\ref{stuffle}) are written 
for a given weight $w=\sum a_i+\sum b_i$ and then are translated into
the language of MPL $G(a_1,\ldots,a_w;1)$.
The relations are written for all the cases where both factors are convergent:
one excludes the cases $a_{w_{1}}=1, p_{w_{1}}=0$ and $b_{w_{2}}=1, q_{w_{2}}=0$.

The regularization relations \cite{Zhao} that we use are
\be
\left\{L_1(0) L_{a_1,\ldots,a_{k}}(q_1,\ldots,q_{k})\right\}_{\rm stuffle}
- \left\{L_1(0) L_{a_1,\ldots,a_{k}}(q_1,\ldots,q_{k})\right\}_{\rm shuffle}
=0
\;,
\label{regularization}
\ee
where the first term in the difference is given by the right-hand side of  (\ref{stuffle})
and the second term is obtained by writing down both factors in terms of MPL and
applying the shuffle relation (\ref{shuffle}). We consider
$q_k \neq 0$ in the case when $a_k = 1$.
The numbers $L_1(0)$ correspond to the variable $T$ introduced in \cite{Zagier}
in the case of multiple zeta values
and used in \cite{Zhao}
in the case of MPL at $n$-th roots of unity. These quantities are infinite and a
regularization is implied, e.g. the upper limit in sums like in \cite{Zagier}. 
Singular terms are cancelled in the difference.
Relations (\ref{regularization}) are nothing but a part of regularized double shuffle relations
of \cite{Zagier,Zhao}.  

If we write down regularized double shuffle relations \cite{Zhao}
in the case, where at most the first power of the variable $T$ is present,
and translate them into the language of MPL we obtain just the
regularization relations (\ref{regularization}).
In the case of multiple zeta values, these first order in $T$ relations are
sufficient and relations with more powers of $T$ do not provide additional
information \cite{Zagier}. This phenomenon takes place also in our case: if we
write down more general regularization relations, i.e. with
$L_{1,\ldots,1}(0,\ldots,0)$ instead of $L_1(0)$, we do not obtain additional
independent relations.

The distribution relations take the form
\be
{\rm Li}_{a_1,\ldots,a_{k}}(x_1,\ldots,x_{k})=d^{a_1+\ldots+a_k-k}\sum_{(y_1,\ldots,y_{k}):\,y_j^d=x_j, j=1,\ldots,k}
{\rm Li}_{a_1,\ldots,a_{k}}(y_1,\ldots,y_{k})
\label{distribution}
\ee
for $d=2,3,6$, where it is assumed that $x_i$ are all $(6/d)$-th roots of unity. We consider all possible
distribution relations, i.e. where only convergent quantities are involved.
 
We write down the shuffle, stuffle, regularization and distribution relations,
then turn to the real and imaginary parts of the MPL according to (\ref{reim})
and use also the complex conjugation relations
\be
G(a_1^*,\ldots,a_{w}^*;1)=G(a_1,\ldots,a_{w};1)^*
\label{cc}
\ee
where we have $r_1^*=r_2,r_3^*=r_4$. 

The total number of these five sets of relations grows fast when the weight is increased.
At weight 6, we have
654452 equations for the real parts and 654937 equations for the imaginary parts.
To construct bases $B_R(w)$ and $B_I(w)$ in $M_R(w)$ and $M_I(w)$, correspondingly, we solve these relations.
 
We construct bases up to weight 6 recursively with the respect to the weight.
At the same time we derive explicit formulae which linearly express 
(with rational coefficients)  all the other elements in terms of the elements of the bases.
For a given weight $w$, let us assume that we have already constructed
bases $B_{R/I}(w')$ for all $w'<w$.
Thus, we have explicit formulae which linearly express 
any given element from $M_R(w')$ and $M_I(w')$ in terms of elements of 
the bases $B_R(w')$ and $B_I(w')$. 
 
Then we proceed at weight $w$. First, we include into $B_R(w)$ and $B_I(w)$
products of the elements of the bases of lower weights, i.e. from various 
$B_R(w')$ and $B_I(w')$. All the other products of the spaces $M_R(w)$ and $M_I(w)$
can be linearly expressed in terms of these products.
Then we continue constructing the bases $B_{R/I}(w)$ by including elements of the type $G_{R/I}(a_1,\ldots,a_w)$.
To do this, we solve the five sets of the relations, where 
unknowns are the $6\times 7^{w-1}$ numbers $G_R(a_1,\ldots,a_w)$ or $G_I(a_1,\ldots,a_w)$.

In this recursive setup, we do not need lifted relations obtained by multiplying a relation among MPL of weight $w'$ by 
an MPL of weight $w-w'$ and applying then shuffle or stuffle relations for resulting products.
This is because the information about 
all the relations of weights $w'=1,2\ldots,w-1$ was already taken into account
and is encoded in the explicit solutions, while shuffle or stuffle relations of weight $w$
are taken into account at the current step.
The relations for the $6\times 7^{w-1}$ numbers $G_R(a_1,\ldots,a_w)$ or $G_I(a_1,\ldots,a_w)$
are linear equations. We solve them for $w=1,2,\ldots,6$ with a code written in  {\tt Mathematica}.
Then we include the numbers on the right-hand sides of these solutions into
$B_R(w)$ and $B_I(w)$.
At the same time, we obtain explicit formulae which linearly express all  
the $G_R(a_1,\ldots,a_w)$ and $G_I(a_1,\ldots,a_w)$ in terms of the elements of 
these $B_R(w)$ and $B_I(w)$. Let us denote by $N_{R/I}(w)$ the numbers of elements of
$B_{R/I}(w)$ of the type $G_{R/I}(a_1,\ldots,a_w)$.  

Zhao conjectured (in Conjecture 9.3 of \cite{Zhao}) that whenever 
$N$ is {\em non-standard}, i.e. has at least two prime factors (e.g. $N=6$ in our case)
the various relations under consideration are not sufficient to reveal bases of
the spaces of MPL at $N$th roots of unity.
Indeed, we discovered that the resulting constants, independent in the sense of
these relations, are still $\mathbb{Q}$-linearly dependent,  
i.e. one can  linearly express some of them in terms of a smaller set of the constants and products of
constants of lower weights. We did this with experimental mathematics using the 
{\tt PSLQ} algorithm \cite{PSLQ} as follows. After we discovered that a given set of elements present on the right-hand sides
of our current solutions is linearly dependent, we started from the set of 
products of constants of lower weights which in all the cases was linearly independent. 
Then we added one by one elements of the given weight (present on the right-hand sides of the solutions) and, 
at each step, checked via {\tt PSLQ} whether
such an element can be expressed in terms of the previously collected elements or not and, if
it could not be reduced to the previous set of elements, we included it into this set.
For the evaluation of MPL, with the accuracy up to 4000 digits we
applied the computer implementation \cite{GiNaC_code} of {\tt GiNaC} \cite{GiNaC}.

In the following table, we show dimensions $D_{R/I}(w)$  
of our bases and the numbers $N_{R/I}(w)$ defined above, i.e. the numbers of algebra generators.
In the last two columns, we show how many elements were
deleted from the primary versions of the bases (obtained by solving linear equations),
using our strategy based on the {\tt PSLQ} algorithm, when arriving at the bases 
$D_{R/I}(w)$.
 
\vspace*{2mm}
\begin{tabular}{|c|c|c|c|c|c|c|}
\hline
w        &  $N_{R}(w)$ &  $D_R(w)$  & $N_{I}(w)$&  $D_I(w)$ & $PSLQ_R$ & $PSLQ_I$\\
\hline
1&              2&                 2&          1&       1 & &\\
2&              1&                 5&          1&       3 & &\\
3&              3&                 12&         2&       9 & &2\\
4&              5&                 30&         5&       25 &2 &6\\
5&              13&                76&         11&      68 &11 &17\\
6&              25&                195&        25&      182 &39 &49\\
\hline
\end{tabular}
\vspace*{2mm}

Elements of the type $G_{R/I}(a_1,\ldots,a_w)$ of our bases up to weight 6 are presented in the Appendix. 
Moreover, bases up to weight~3 are explicitly expressed there in terms of known numbers, following~\cite{Broadhurst:1998rz}.
Files with the complete bases and files with reductions of any MPL up to weight 6 to the elements of the bases 
can be downloaded from \url{http://theory.sinp.msu.ru/~smirnov/mpl6}.
The files with reductions {\tt sl1re.m,sl1im.m,}, \dots {\tt sl6re.m}, {\tt sl6im.m} are 
in the text format and contain lists of substitutions which give results 
for all possible MPL under consideration as linear combinations of the elements of our bases.
The notation in the files exactly follows the notation used in the paper, i.e.
{\tt GR[a[1],\ldots,a[w]]} and {\tt GI[a[1],\ldots,a[w]]}
stand for $G_{R/I}(a_{[1]},\ldots,a_{[w]})$.  
The files with the elements of the bases are {\tt irRe.m} and {\tt irIm.m}. After loading
them, one obtains {\tt irRe[1]},\ldots,{\tt irIm[6]} as the elements of the bases.

The files with reductions to the bases for weights~5 and~6 are rather big. We would recommend
to use them on a computer with at least 10 gb RAM.
For convenience, we also present files with 4000 digits results for the elements of the bases.

\section{Discussion and conclusion}

Broadhurst found \cite{Broadhurst_lecture} that the dimensions of the spaces
$M_{R/I}(w)$ are given by $D_{R/I}(w)=(F_{2w+2} \pm F_{w+1})/2$, where $F_n$ is a Fibonacci number. 
Our results up to weight six are in agreement with this result, i.e. the numbers $D_{R/I}(w)$
of the elements of $B_{R/I}(w)$ are given by these formulae.

Broadhurst suggested~\cite{Broadhurst_lecture}, at any $w$, conjectured bases of $M_R(w)$  and $M_I(w)$,
using only letters $0,-1,r_4$. 
For weight~6, the elements of the bases are, in our notation,  
\bea
\{G(0, 0, 0, 0, r_4, r_4), G(0, 0, 0, 0, r_4, -1), G(0, 0, 0, r_4, 0, r_4), 
&& \nn \\  && \hspace*{-107mm} 
G(0, 0, r_4, r_4, r_4, r_4), 
G(0, 0, r_4, r_4, r_4, -1),  G(0, 0, r_4, r_4, -1, r_4), 
\nn \\  && \hspace*{-107mm}
G(0, 0, r_4, r_4, -1, -1), 
 G(0, 0, r_4, -1, r_4, r_4), 
 G(0, 0, r_4, -1, r_4, -1),
\nn \\  && \hspace*{-107mm} 
 G(0, 0, r_4, -1, -1, r_4), 
 G(0, 0, r_4, -1, -1, -1), 
 G(0, r_4, 0, r_4, r_4, r_4),
\nn \\  && \hspace*{-107mm}
 G(0, r_4, 0, r_4, r_4, -1),
 G(0, r_4, 0, r_4, -1, r_4), G(0, r_4, 0, r_4, -1, -1), 
\nn \\  && \hspace*{-107mm}  
 G(0, r_4, r_4, 0, r_4, -1), 
 G(r_4, r_4, r_4, r_4, r_4, -1), 
 G(r_4, r_4, r_4, r_4, -1, -1), 
\nn \\  && \hspace*{-107mm}  
 G(r_4, r_4, r_4, -1, r_4, -1), 
G(r_4, r_4, r_4, -1, -1, -1), 
G(r_4, r_4, -1, r_4, -1, -1), 
\nn \\  && \hspace*{-107mm} 
G(r_4, r_4, -1, -1, r_4, -1), 
G(r_4, r_4, -1, -1, -1, -1), 
\nn \\  && \hspace*{-107mm} 
 G(r_4, -1, r_4, -1, -1, -1), G(r_4, -1, -1, -1, -1, -1)\} \;,
\label{BB}
\eea
where $G$ stands either for $G_R$ or $G_I$.
Our results up to weight six confirm this conjecture. To see this,
we reduced the elements of (\ref{BB}) to our bases
and found that they are connected by invertible matrices.
Our results up to weight 6 also confirm the generalized parity conjecture\footnote{
This conjecture has been quite recently proven in a much more general case -- see \cite{Panzer:2016}.}
by Broadhurst~\cite{Broadhurst:1998rz}.
  
Our results are based on the basis which appeared in the process of solving
multiple linear equations and we did not try to turn to ``better'' bases
connected with ours by linear transformations.
For practical applications in high-energy physics, our basis looks quite sufficient.
We have already applied the results presented in this paper in
the evaluation of three-loop massive form factors \cite{Henn:2016kjz} and 
the corresponding master integrals \cite{Henn:2016tyf}.
Of course, we replaced lower-weight constants in terms of known constants, like
this is done up to weight 3 in eqs.~(\ref{w3}).
Since MPL with the alphabet of eq. (\ref{letters}) naturally appear in many 
physically important problems, our results will also be useful in future
calculations. 

We understand that our bases are not mathematically natural and that there
are various better choices. It looks like the most elegant choice 
would be to turn to the bases conjectured by Broadhurst~\cite{Broadhurst_lecture}
and consisting only letters $0,-1,r_4$. Another natural choice would be
to include into the bases as many elements with the letters $0,-1,1$
(i.e. alternating Euler sums) as possible, i.e. to use the bases described in~\cite{Broadhurst:1996az} as subbases.
Still if one is interested only in elements with the letters $0,-1,1$,
i.e. to understand what $G(a_1,\ldots,a_6;1)$  is, with $a_i$ taken from this
subalphabet one can apply the package {\tt HPL} \cite{Maitre:2005uu} and immediately
obtain a result in terms of known numbers.
Similarly, one can include bases with letters $0,1,r_1$ \cite{Broadhurst:2014jda}.
However, the task of turning from our basis to some other basis does not 
look complicated. Although manipulating with our database requires big RAM,
a reordering of not more than 200 elements by expressing some elements 
in terms of ``better'' elements using our results is rather straightforward.

According to the Zhao's conjecture \cite{Zhao} mentioned above,
the number of MPL at $N$th roots of unity coincides
with the upper bounds of the motivic theory on ${\mathbb P}^1\backslash\{0,\infty, r_{i} \}$. 
These upper bounds for the motivic fundamental group were computed by
Deligne and Goncharov  \cite{DG} (in Section~5). 
Moreover, the numbers $D(w)$  of the dimension of the vector space $M(w)$
explicitly given in \cite{DG} by coefficients in
\be
\sum_w D(w) t^w=\frac{1}{1-3t+t^2}=1 + 3 t + 8 t^2 + 21 t^3 + 55 t^4 + 144 t^5 + 377 t^6 +O(t^7)
\nn
\ee
exactly coincide at $N=6$ with our dimensions $D(w)=D_R(w)+D_I(w)$ obtained by solving the various relations
between MPL {\em and} revealing extra relations with {\tt PSLQ}.
 
Since our results were obtained with the help of the {\tt PSLQ} algorithm
they have the status of experimental mathematics.
However, we find them rather convincing: the missing relations among 
candidate elements for the bases were revealed with {\tt PSLQ} using the accuracy of up to 4000 digits,
while the final checks were made with 4300 digits.
Nevertheless, it would be very interesting to confirm the part of the relations
found here with PSLQ, e.g., using the methods outlined in refs.
\cite{Brown:2011ik,Panzer:2016snt}. 
However, this is beyond the scope of the present paper.
Also beyond the scope of this paper, a natural next
step would be to try to express all the elements of our bases in terms of
the Clausen function, polylogarithms etc., similarly to how this was done \cite{Broadhurst:1998rz} up to weight~3.

\vspace{0.2 cm}
{\em Acknowledgments.}
We are grateful to David Broadhurst, Francesco Moriello and Eric Panzer for fruitful discussions
and various pieces of advice.
 J.M.H. is supported in part by a GFK fellowship and by
the PRISMA cluster of excellence at Mainz university.
The work of A.S. and V.S. is supported by RFBR, grant 17-02-00175A.
The work of V.S. was partially
supported by the Alexander von Humboldt Foundation (Humboldt Forschungspreis). 

\appendix

\section{Bases}

Here we describe, for $w=1,\ldots,6$, the elements of our bases of the
form $G_{R/I}(a_1,\ldots,a_w)$.

At weight one, our basis for the real parts of $G(a;1)$  contains two elements
$\{G_R(-1), G_R(r_4)\}$
and our basis for the imaginary parts of $G(a;1)$ contains one element $G_I(r_2)$.
At weight two, our basis for the real parts of $G(a_1,a_2;1)$  contains element
$G_R(r_2, -1)$
and our basis for the imaginary parts of $G(a_1,a_2;1)$ contains $G_I(0, r_2)$.
At weight three, our basis for the real parts of 
$G(a_1,a_2,a_3;1)$  contains 3 elements
$\{
G_R(0, 0, 1), G_R(r_2, 1, -1), G_R(r_2, 1, r_3)
 \}$
and our basis for the imaginary parts of 
$G(a_1,a_2,a_3;1)$  contains 2 elements
$\{
 G_I(0, 1, r_4), G_I(0, r_2, -1) 
  \}$.

These elements are expressed in terms of the constants revealed in~\cite{Broadhurst:1998rz}
as follows:
\bea
G_R(-1)&=& \log (2)\;, \nn \\
G_R(r_4)&=& \frac{1}{2}\log(3)\;, \nn \\
G_R(r_2,-1)&=& \frac{1}{4} \text{Li}_2\left(\frac{1}{4}\right)
+\frac{\pi^2}{72}+\frac{1}{2}\log ^2(2)-\frac{1}{2} \log (2) \log (3)\;, \nn \\
G_R(0,0,1)&=& -\zeta(3)\;, \nn \\
G_R(r_2,1,-1))&=& -\frac{5 \pi  }{18}{\rm Cl}_2\left(\frac{\pi}{3}\right)
+2{\rm Re}\left(\text{Li}_3\left(\frac{r_1}{2}\right)\right)
+\frac{1}{4}\text{Li}_2\left(\frac{1}{4}\right) \log (2) \nn\\
&&+\frac{17 }{72}\zeta (3)
+\frac{1}{6}\log^3(2)-\frac{1}{72} \pi ^2 \log (2)\;,\nn\\
G_R(r_2,1,r_3)&=& -\frac{7 \pi }{36}{\rm Cl}_2\left(\frac{\pi}{3}\right)
+2 {\rm Re}\left(\text{Li}_3\left(\frac{i}{\sqrt{3}}\right)\right)
-\frac{1}{8}\text{Li}_2\left(\frac{1}{4}\right) \log (3)
\nn \\ &&
+\frac{25 \zeta (3)}{36}-\frac{\log^3(3)}{24}+\frac{1}{4} \log (2) \log ^2(3)
\nn \\ &&
-\frac{1}{4} \log ^2(2) \log (3)
-\frac{7}{144} \pi ^2 \log (3)\;, \nn\\
G_I(r_2)&=& -\frac{\pi }{3}\;, \nn \\
G_I(0,r_2)&=&-{\rm Cl}_2\left(\frac{\pi}{3}\right)\;, \nn \\
G_I(0,1,r_4)&=& -\frac{1}{3} {\rm Cl}_2\left(\frac{\pi}{3}\right) \log (3)
-\frac{8}{5} {\rm Im}\left(\text{Li}_3\left(\frac{i}{\sqrt{3}}\right)\right)
\nn \\ &&
+\frac{41 \pi ^3}{810}+\frac{1}{30} \pi \log ^2(3)\;, \nn \\
G_I(0,r_2,-1)&=& -\frac{5}{3} {\rm Cl}_2\left(\frac{\pi}{3}\right) \log (2)
+\frac{107 \pi ^3}{3240}
-2 {\rm Im}\left(\text{Li}_3\left(\frac{r_1}{2}\right)\right)
\nn \\ &&
+\frac{6}{5} {\rm Im}\left(\text{Li}_3\left(\frac{i}{\sqrt{3}}\right)\right)
-\frac{1}{40} \pi  \log ^2(3)+\frac{1}{6} \pi  \log ^2(2) \;,
\label{w3}
\eea
where ${\rm Cl}_2\left(\theta\right)=\sum_{k=1}^{\infty} \sin(k \theta)/k^2$.
  
At weight four, our basis for the real parts of 
$G(a_1,\ldots,a_{4};1)$  contains 5 elements
$\{G_R(0, 0, r_2, -1), G_R(0, 0, r_4, 1), G_R(r_2, 1, 1, -1),$ 

\noindent $G_R(r_2, 1, 1, r_3), G_R(r_2, 1, r_2, -1)\}$
and our basis for the imaginary parts of 
$G(a_1,\ldots,a_{4};1)$  contains 5 elements
$\{G_I(0, 0, 0, r_2), $

\noindent $G_I(0, 1, 1, r_4), G_I(0, 1, r_2, -1), G_I(0, 1, r_2, r_3),
  G_I(0, r_2, 1, -1)\}\;.$

At weight five, our basis for the real parts of 
$G(a_1,\ldots,a_{5};1)$  contains 13 elements
\bea
\{G_R(0, 0, 0, 0, 1), G_R(0, 0, 1, 1, -1), G_R(0, 0, 1, 1, r_4), 
 G_R(0, 0, 1, r_2, -1),  
  \nn \\  && \hspace*{-130mm} 
G_R(0, 0, 1, r_2, r_3), G_R(0, 0, 1, r_2, r_4), 
 G_R(0, 0, r_2, 1, -1), 
 G_R(r_2, 1, 1, -1, r_2), 
  \nn \\  && \hspace*{-130mm} 
G_R(r_2, 1, 1, 1, -1),  G_R(r_2, 1, 1, 1, r_3), G_R(r_2, 1, 1, r_2, -1), 
 \nn \\  && \hspace*{-130mm} 
G_R(r_2, 1, 1, r_2, r_3),  G_R(r_2, 1, 1, r_4, -1) 
 \} 
 \nn
\eea
and our basis for the imaginary parts of 
$G(a_1,\ldots,a_{5};1)$  contains 11 elements
\bea
\{G_I(0, 0, 0, 1, r_2), G_I(0, 0, 0, 1, r_4), G_I(0, 0, 0, r_2, -1), 
 G_I(0, 1, 1, -1, r_2), 
  \nn \\  && \hspace*{-130mm} 
G_I(0, 1, 1, -1, r_4),  G_I(0, 1, 1, 1, r_4), 
 G_I(0, 1, 1, r_2, r_3), 
 G_I(0, 1, 1, r_4, -1), 
 \nn \\  && \hspace*{-130mm} 
G_I(0, 1, 1, r_4, r_1),  G_I(0, 1, r_2, r_3, r_2), G_I(0, r_2, 1, 1, -1)\}\;.
 \nn
\eea

At weight six, our basis for the real parts of 
$G(a_1,\ldots,a_{6};1)$  contains 25 elements
\bea
\{G_R(0, 0, 0, 0, r_2, -1), G_R(0, 0, 0, 0, r_2, r_4), 
 G_R(0, 0, 0, 0, r_4, 1), 
  \nn \\  && \hspace*{-114mm} 
G_R(0, 0, 1, 1, -1, r_2),  G_R(0, 0, 1, 1, -1, r_4), G_R(0, 0, 1, 1, 1, -1), 
   \nn \\  && \hspace*{-114mm}  
 G_R(0, 0, 1, 1, 1, r_4),
  G_R(0, 0, 1, 1, r_2, -1),   G_R(0, 0, 1, 1, r_2, r_3), 
   \nn \\  && \hspace*{-114mm} 
 G_R(0, 0, 1, 1, r_4, -1), G_R(0, 0, 1, 1, r_4, r_1), 
 G_R(0, 0, 1, 1, r_4, r_2), 
  \nn \\  && \hspace*{-114mm} 
 G_R(0, 0, 1, r_2, 1, -1), 
 G_R(0, 0, 1, r_2, 1, r_3), G_R(0, 0, 1, r_2, r_1, -1), 
 \nn \\  && \hspace*{-114mm} 
 G_R(0, 0, 1, r_2, r_1, r_4), 
 G_R(r_2, 1, 1, 1, -1, r_2), 
 G_R(r_2, 1, 1, 1, -1, r_3), 
\nn \\  && \hspace*{-114mm} 
 G_R(r_2, 1, 1, 1, 1, -1), 
 G_R(r_2, 1, 1, 1, 1, r_3),  G_R(r_2, 1, 1, 1, r_2, -1), 
  \nn \\  && \hspace*{-114mm} 
G_R(r_2, 1, 1, 1, r_2, r_3), G_R(r_2, 1, 1, 1, r_3, r_2), 
  \nn \\  && \hspace*{-114mm} 
 G_R(r_2, 1, 1, 1, r_4, -1), G_R(r_2, 1, 1, r_2, -1, r_2)\}
 \nn
\eea
and our basis for the imaginary parts of 
$G(a_1,\ldots,a_{6};1)$  contains 25 elements
\bea
\{G_I(0, 0, 0, -1, 1, r_2), G_I(0, 0, 0, -1, 1, r_4), 
 G_I(0, 0, 0, -1, r_2, 1), 
\nn \\  && \hspace*{-116mm}
G_I(0, 0, 0, 0, 0, r_2), G_I(0, 0, 0, 1, 1, r_4),
  G_I(0, 0, 0, 1, r_2, -1), 
\nn \\  && \hspace*{-116mm}  
  G_I(0, 0, 0, r_2, 1, -1), 
 G_I(0, 0, 0, r_2, 1, r_3), G_I(0, 0, 0, r_2, r_2, -1),
\nn \\  && \hspace*{-116mm}
 G_I(0, 1, 1, -1, r_2, -1), G_I(0, 1, 1, -1, r_2, r_1), 
 G_I(0, 1, 1, -1, r_2, r_4), 
 \nn \\  && \hspace*{-116mm}
 G_I(0, 1, 1, -1, r_4, r_3), 
 G_I(0, 1, 1, 1, -1, r_2), G_I(0, 1, 1, 1, -1, r_4), 
 \nn \\  && \hspace*{-116mm} 
 G_I(0, 1, 1, 1, 1, r_4),  G_I(0, 1, 1, 1, r_2, r_3), 
 G_I(0, 1, 1, 1, r_4, -1),
 \nn \\  && \hspace*{-116mm}
 G_I(0, 1, 1, 1, r_4, r_1), 
 G_I(0, 1, 1, r_2, -1, r_2),  G_I(0, 1, 1, r_2, -1, r_3), 
 \nn \\  && \hspace*{-116mm}
 G_I(0, 1, 1, r_2, -1, r_4), G_I(0, 1, 1, r_2, r_3, -1), 
 \nn \\  && \hspace*{-116mm}
 G_I(0, r_2, 1, 1, -1, r_2), G_I(0, r_2, 1, 1, 1, -1)\}\;.
\nn  
\eea

\bibliographystyle{JHEP}


\end{document}